\documentclass[prd,twocolumn,showpacs,preprintnumbers,amsmath,amssymb]{revtex4}

\usepackage{graphicx}

\usepackage{bm}

\newcommand{\beq}{\begin{equation}}
\newcommand{\eeq}{\end{equation}}
\newcommand{\beqs}{\begin{eqnarray}}
\newcommand{\eeqs}{\end{eqnarray}}
\newcommand{\lsim}{\mathrel{\raisebox{-
.6ex}{$\stackrel{\textstyle<}{\sim}$}}}

\newcommand{\pslash}{p\hspace{-0.067in}\slash}

\begin{document}

\title{Implications of Dynamical Generation of Standard-Model Fermion Masses}

\author{Neil D. Christensen}

\author{Robert Shrock}

\affiliation{
C.N. Yang Institute for Theoretical Physics \\
State University of New York \\
Stony Brook, NY 11794}

\begin{abstract}

We point out that if quark and lepton masses arise dynamically, then in a wide
class of theories the corresponding running masses $m_{f_j}(p)$ exhibit the
power-law decay $m_{f_j}(p) \propto \Lambda_j^2/p^2$ for Euclidean momenta $p
\gg \Lambda_j$, where $f_j$ is a fermion of generation $j$, and $\Lambda_j$ is
the maximal scale relevant for the origin of $m_{f_j}$.  We estimate resultant
changes in precision electroweak quantities and compare with current data.  It
is found that this data allows the presence of such corrections.  We also note
that this power-law decay renders primitively divergent fermion mass
corrections finite.

\end{abstract}

\maketitle

The origin of fermion masses in the standard model (SM) remains mysterious.
This model accomodates quark and charged lepton masses via Yukawa couplings to
a postulated Higgs boson, but this does not provide insight into these masses,
especially since it requires small dimensionless Yukawa couplings for all of
the fermions except the top quark, ranging down to $10^{-6} - 10^{-5}$ for the
first generation, with no explanation.  This suggests that one consider models
in which quark and lepton masses arise dynamically.  The standard model also
predicted zero neutrino masses, and has had to be augmented (with both Dirac
and Majorana bilinears, in general) to incorporate the observed nonzero
neutrino masses.  Early studies on dynamically generated standard-model fermion
masses found that at asymptotically large Euclidean momenta $p$ they would
decay like a power of $p$ \cite{jbw}-\cite{el}.  We point out here that in a
wide class of theories in which quark and lepton masses arise dynamically, the
corresponding running masses of the fermions in the $j$'th generation exhibit
the power-law decay $m_{f_j}(p) \propto \Lambda_j^2/p^2$ for Euclidean $p$
large compared to the largest scale $\Lambda_j$ involved in this mass
generation.  This renders proper fermion self-energy corrections finite, rather
than divergent.  An important question for these theories is whether such
power-law decays of SM fermion masses are allowed by current precision data.
We address this question here and answer it in the affirmative.

We consider a class of theories in which (i) the physics underlying the
generation of lepton and (current) quark masses involves the formation of a set
of bilinear electroweak symmetry-breaking (EWSB) fermion condensates, denoted
generically as $\{\langle \bar \psi \psi \rangle \}$ at a given scale
$\Lambda_{EW}$, (ii) there are interactions connecting the $\psi$'s with SM
fermions to communicate the EWSB to these fermions, and (iii) the interactions
responsible for this mass generation are asymptotically free, so that at
sufficiently large momenta all couplings are small, and hence $\langle \bar
\psi \psi \rangle$ and other relevant operators have essentially canonical
dimensions with small anomalous dimensions.

We begin by showing the power-law decay of the running mass of an SM fermion in
two classes of models satisfying our premises above and then give a general
argument. First, consider extended technicolor (ETC) models
\cite{el,etc2,etcrev} containing a subsector with a set of massless fermions
$\{F\}$ subject to an asymptotically free, vectorial, confining gauge
interaction denoted technicolor (TC) \cite{tc,dsb}.  As the momentum decreases
from high values to the scale $\Lambda_{TC} = \Lambda_{EW} \sim 250$ GeV, the
TC gauge coupling grows sufficiently to produce EWSB technifermion condensates
$\langle \bar F F \rangle \equiv \langle \bar \psi \psi \rangle$.  The theory
includes a set of vector bosons that transform SM fermions into technifermions,
thereby communicating the EWSB to the SM fermions and dynamically generating
masses for them.  These ETC gauge bosons are denoted $V^j_t$, where $j=1,2,3$
is a generation (family) index and $t$ is a technicolor index.  The ETC group
is an asymptotically free, anomaly-free, chiral gauge theory that breaks in
stages at the scales $\Lambda_j$, to the residual exact TC gauge group.
Typical values used in recent studies are $\Lambda_1 \simeq 10^3$ TeV,
$\Lambda_2 \simeq 10^2$ TeV, and $\Lambda_3 \simeq 4$ TeV
\cite{at94}-\cite{kt}.  This hierarchy in breaking scales produces a hierarchy
in the resultant quark and lepton masses, which have inverse power dependence
on these scales.  Although fully realistic ETC models have not yet been
constructed, and they are subject to strong constraints such as those from
precision electroweak data, they serve as useful explicit examples of dynamical
fermion mass generation.

To investigate the dependence of the running mass $m_{f_j}(p)$ on momentum,
consider the one-loop diagram in which $f_j(p)$ emits a virtual $V^j_t$ with
momentum $p-k$ and mass $M_j \simeq \Lambda_j$, transforming to a technifermion
$F^t$ which reabsorbs the $V^j_t$.  We neglect fermion mixing to start.  After
Wick rotation, this diagram yields
\beqs
& & m_{f_j}(p) \sim \cr\cr
& & g_{_{ETC}}^2 N_{TC} \int \frac{d^4 k}{(2\pi)^4} \
\frac{\Sigma_{TC}(k)}{[k^2+\Sigma_{TC}(k)^2][(p-k)^2+M_j^2]} \ ,  \cr\cr
& & 
\label{integraltc}
\eeqs
where $\Sigma_{TC}(k)$ is the dynamical technifermion mass \cite{gi}, with
$\Sigma_{TC}(k) = \Sigma_{TC,0} \simeq 2 \Lambda_{TC}$ for Euclidean $k \ll
\Lambda_{TC}$.  In early TC theories, $\Sigma_{TC}(k)$ had the power-law decay
$\Sigma_{TC} \simeq \Sigma_{TC,0}/k^2$ for $k^2 \gg \Lambda_{TC}^2$, analogous
to the momentum dependence of the constituent quark mass in quantum
chromodynamics (QCD) \cite{sig}, with $\Lambda_{QCD}$ replaced by
$\Lambda_{TC}$.  Current TC theories rely upon a TC gauge coupling that runs
slowly (walks) \cite{wtc} in the interval $\Lambda_{TC} \lsim k \lsim
\Lambda_w$, where typically $\Lambda_w \simeq \Lambda_3$.  In walking TC
theories, $\Sigma_{TC}(k)$ falls like $\Sigma_{TC,0}^2/k$ for $\Lambda_{TC}
\lsim k \lsim \Lambda_w$ and like $\Sigma_{TC,0}^3/k^2$ for larger $k$. This
yields the pole mass $m_{f_j} \sim \kappa \eta \Lambda_{TC}^3/\Lambda_j^2$,
where $\kappa \sim O(10)$ is a numerical factor (see, e.g., \cite{ckm}) and
$\eta$ is a walking factor \cite{eta}.  The walking can result from
an approximate infrared fixed point in the TC theory; it enhances SM
fermion masses, raises pseudo-Nambu-Goldstone boson masses, and can reduce TC
contributions to the electroweak $S$ parameter \cite{scalc}.

Expanding eq. (\ref{integraltc}) for small $p$ and using $M_j^2 \gg
\Sigma_{TC,0}^2$, we find
\beq
m_{f_j}(p) \simeq m_{f_j}\Bigl ( 1 - a_{sp}\frac{p^2}{\Lambda_j^2} \Bigr )  
\label{mfsmallp}
\eeq
for $p^2 \ll \Lambda_j^2$, where $a_{sp}$ is a positive constant $\sim O(1)$
depending on the dynamics responsible for the fermion masses, and we neglect
possible $p$-dependent log factors. From eq. (\ref{mfsmallp}), it is evident
that softness effects for $m_{f_j}$ become significant for $p \sim \Lambda_j$. 

Expanding eq. (\ref{integraltc}) for $p \gg \Lambda_j$, we find
\beq
m_{f_j}(p) \simeq m_{f_j}\frac{\Lambda_j^2}{p^2} \ , 
\label{mflargep}
\eeq
Here we neglect factors $(\ln(p^2/\mu^2))^{a_f}$ arising from the anomalous
dimensions of the full operator $\bar f_j f_k \bar\psi\psi$, since these are
subdominant compared with the power-law decay.  Since this $p$-dependence holds
above the walking interval $\Lambda_{TC} \lsim p \lsim \Lambda_3$, it is
essentially independent of the presence of walking, the effect of which is
contained in $\Sigma_{TC}(k)$ and the pole mass, $m_{f_j}$.  Hence, one can
verify that this type of model yields eq. (\ref{mflargep}) by inserting an
explicit approximate functional form such as $\Sigma_{TC}(k)
=\Sigma_{TC,0}/(1+k^2/\Lambda_{TC}^2)$ in eq. (\ref{integraltc}).  Because the
TC theory is strongly coupled, higher-loop diagrams are also important; these
are subsumed in the low-energy effective Lagrangian containing four-fermion
terms of the form $\bar f_j f_k \bar F F$.  Given our premises, these do not
significantly modify eq. (\ref{mflargep}).  As indicated, mixing terms can be
included, as in Refs. \cite{nt,ckm,kt}.  While it is challenging to construct
ETC models leading to fully realistic quark mixing, plausible models with small
off-diagonal entries in $M^{(u)}_{jk}$ and $M^{(d)}_{jk}$ maintain the
generational dependence of the falloff in eq. (\ref{mflargep}).

A second example is provided by topcolor-assisted technicolor (TC2) models
\cite{tc2,etcrev}.  For these the set of $\langle \bar\psi\psi\rangle$ includes
both technifermion condensates and condensates directly involving the $t_L$ and
$t_R$ fields, which contribute importantly to both EWSB and the top quark mass.
The relevant scale $\Lambda_t$ for the latter condensate(s) is of order a TeV,
$\lsim \Lambda_3$. Since theories of this type can satisfy our premise, it
follows that eqs. (\ref{mfsmallp}) and (\ref{mflargep}) hold.

The hierarchy of scales $\Lambda_3 < \Lambda_2 < \Lambda_1$ means that the
power-law decay of $j=3$ fermion masses sets in at momenta where the mass
matrix elements $M^{(f)}_{jk}$ with $j,k \in \{1,2\}$, are, up to logs, still
largely momentum-independent (hard).  Up to small mixing effects, as $p$
increases above $\Lambda_3$, the $t$, $b$, and $\tau$ running masses begin to
decay as in (\ref{mflargep}) with $j=3$; then as $p$ increases above
$\Lambda_2$, the $c$, $s$, and $\mu$ masses fall off like $\Lambda_2^2/p^2$,
and finally, for $p  > \Lambda_1$, the $u$, $d$, and
$e$ masses fall off like $\Lambda_1^2/p^2$. 

We next give a general analysis of the behavior of $m_{f_j}(p)$ in a theory
with dynamical fermion mass generation satisfying our premises.  The relevant
low-energy effective Lagrangian contains the terms
\beq
{\cal L}_{eff} = \sum_{f,j,k,\psi} 
b^{(f)}_{jk}\bar f_{jL} f_{kR}\bar \psi \psi + h.c. \ , 
\label{leff}
\eeq
where $f$ denotes a quark or charged lepton, $j,k \in \{1,2,3\}$ are generation
indices, and we suppress other terms involving Fierz rearrangements \cite{njl}.
The existence of the condensate(s) $\langle \bar \psi \psi\rangle$ yields the
bilinear mass terms $\bar f_{jL}M^{(f)}_{jk} f_{kR} + h.c.$, where
$M^{(f)}_{jk} = b_{jk}^{(f)}\langle \bar \psi \psi\rangle$.  Diagonalizing this
matrix $M^{(f)}$, one obtains the (physical, pole) masses $m_{f_j}$.  To
account for the generational hierarchy in SM fermion masses, the dynamics
should produce $b^{(f)}_{jk}$'s such that $m_{f_j} = b^{(f)}_{diag.,j} \langle
\bar\psi\psi \rangle = c^{(f)}_j \langle \bar\psi\psi \rangle/\Lambda_j^2$ with
$\Lambda_3 < \Lambda_2 < \Lambda_1$ (where a possible TC2 $\Lambda_t$ is
subsumed as $\simeq \Lambda_3$).  Now, performing the diagonalization at a
Euclidean momentum $p$ to obtain the running masses $m_{f_j}(p)$, and using the
asymptotic freedom property, which guarantees that for large $p$, $\langle
\bar\psi\psi\rangle$ has operator dimension $d-1=3$ with small anomalous
dimensions, it follows that for $p \gg \Lambda_j$, $m_{f_j}(p) \sim c^{(f)}_j
\langle \bar\psi\psi \rangle/p^2$; substituting for $c^{(f)}_j$, one gets
eq. (\ref{mflargep}) \cite{ddim,except}.  The generational dependence of the
power-law decays of running neutrino masses is more complicated, since it
involves both Dirac and Majorana masses of different scales and also
generically large mixing effects, as suggested by the observed large leptonic
mixing angles.

For the theories considered here, although these power-law decays of SM fermion
masses can ultimately be traced to the presence of bilinear fermion
condensate(s), they are distinctively different from the well-studied softness
of constituent quark masses $\Sigma$ in QCD \cite{lane74,sig} and the softness
of dynamical technifermion masses $\Sigma_{TC}$ in either scaled-up QCD-like or
modern walking technicolor theories \cite{gi,wtc}.  In all three of the latter
cases, this softness sets in on the scale of the dynamical mass itself, which
is the scale where the respective QCD or TC gauge interaction gets strong and
breaks the chiral symmetry.  In contrast, the standard-model fermions exhibit
softness on scales which (i) are higher than their pole masses, indeed many
orders of magnitude higher for the first and second generations; (ii) are
associated with the communication of the dynamical EWSB sector to these
fermions, and (iii) have a generational hierarchy.

We now use eq. (\ref{mflargep}) as a new test of theories with dynamical
fermion mass generation.  We ask the question: Is current precision electroweak
data consistent with such power-law decays of SM fermion masses?  One of the
cleanest tests is provided by SM fermion loop corrections to $W$ and $Z$
self-energies. Standard-model contributions including those due to fermions,
are considered to be subtracted in defining the oblique electroweak correction
parameters $S$, $T$, and $U$, so that nonzero values of these parameters
indicate new physics \cite{pdg}, \cite{pt}-\cite{erlang}.  These SM fermion
loop contributions are calculated assuming constant fermion masses.

Consider the parameter $\rho = m_W^2/(m_Z^2 \hat c^2)$ \ \cite{v77}, where
$\hat c^2 = \cos^2 \theta_{W,\overline{MS}} = 1- \hat s^2$.  We focus on the
contribution to $\rho$ of the $(t,b)$ quarks, which is the largest among SM
fermions.  The conventional (one-loop, $1\ell$) result for this is \cite{v77}
$(\Delta \rho )_{tb,hard,1\ell} = N_c G_F f_\rho(m_t^2,m_b^2)/(8 \pi^2
\sqrt{2})$, where $N_c=3$ and $f_\rho(x,y)=x+y-2 xy(x-y)^{-1}\ln(x/y)$.
Numerically, $(\Delta \rho )_{tb,hard,1\ell} \simeq 0.98 \times 10^{-2}$.  In
theories with dynamical generation of SM fermion masses, to leading order in
$m_t/\Lambda_3$, we find from eq. (\ref{mflargep}) that
\beq
(\Delta \rho)_{tb} = (\Delta \rho)_{tb,hard} \biggl [ 1 - a_\rho
 \biggl (\frac{m_t^2}{\Lambda_3^2} \biggr ) \biggr ] \ , 
\label{rhosoft}
\eeq
where $a_\rho$ is a positive coefficient $\sim O(1)$ depending on the dynamics
responsible for the generation of these fermion masses.  The positivity of
$a_\rho$ follows from the fact that the softness of the top quark mass reduces
the violation of the custodial SU(2) symmetry and hence the value of $\Delta
\rho$. In theories (e.g., TC2) with topcolor, corrections of order
$(m_t/\Lambda_t)^2$ could also be important.  Numerically, $(m_t/\Lambda_3)^2 =
0.03 (1 \ {\rm TeV}/\Lambda_3)^2$, so that with $\Lambda_3$ on the few TeV
scale, the fractional reduction of $\Delta \rho_{tb}$ in eq. (\ref{rhosoft}) is
of order $10^{-2} - 10^{-3}$.

In passing, we note that the correction (\ref{rhosoft}) is comparable to some
of the terms, such as those $\propto x_t^2$, where $x_t = G_F m_t^2/(8\pi^2
\sqrt{2})$, in the two-loop contribution to $\Delta \rho_{tb,hard}$
\cite{t2loop}, and to the overall three-loop contribution \cite{t3loop}
(including $\alpha_s^2 x_t$ terms $\sim 10^{-4}$) to $\Delta \rho_{tb,hard}$.

Thus for theories with dynamical generation of SM fermion masses, the
conventional subtraction, using the constant-mass expression, to get $T =
\Delta \rho^{(new)}/\hat\alpha(m_Z^2)$, where $\Delta \rho^{(new)}$ denotes the
change in $\rho$ due to new physics (and $\hat\alpha(m_Z^2)$ is the
electromagnetic coupling), would be a slight oversubtraction.  Using a
definition of $T$ with the subtraction of the actual, rather than the
hard-mass, contribution of $(t,b)$ and correcting for the above oversubtraction
thus slightly shifts the allowed region in $T$ upward, by the amount 
\beq
\Delta T_{tb,soft} \simeq 1.3a_\rho \frac{m_t^2}{\Lambda_3^2} \ . 
\label{tcor}
\eeq

The (1-loop) contribution of the hard-mass $(t,b)$ doublet to the full $S$
parameter before subtraction (bf) is 
\beq 
S^{(bf)}_{tb,hard,1\ell} \simeq \frac{N_c}{12\pi}\left [
-\frac{1}{9} + \frac{7 m_Z^2}{30 m_t^2}-\frac{4}{3}\ln \left
(\frac{m_t}{m_Z}\right)\right ] \ , 
\label{stb}
\eeq
equal to $-0.075$.  Modelling a soft $t$-quark mass approximately by a small
decrease in an effective $m_t$ makes this expression slightly less negative.
As with $T$, defining $S$ with the subtraction of the actual, rather than the
hard-mass, $(t,b)$ contribution and correcting for the difference yields a
small shift downward in the allowed region in $S$, by the amount
\beq
\Delta S_{tb,soft} \simeq -0.1 a_S \frac{m_t^2}{\Lambda_3^2} \ , 
\label{scor}
\eeq
where $a_S$ is a positive constant depending on the interactions responsible
for the dynamical generation of SM fermion masses and expected to be $\sim
O(1)$.  

Current global fits yield allowed regions in $(S,T)$ depending on a reference
value of the SM Higgs mass, $m_{H,ref.}$ \cite{pdg}.  Since theories that
dynamically generate SM fermion masses usually also have dynamical electroweak
symmetry breaking without any fundamental Higgs, it is appropriate to insert an
effective scale of order a TeV for $m_{H,ref.}$. The corresponding allowed
region for $m_{H,ref.}=1$ TeV shown in \cite{pdg} is roughly elliptical, with
semimajor axis having positive slope and central values $(S,T) \simeq (-0.21,
0.15)$.  The uncertainties listed (for a given $m_{H,ref.}$) are $\pm 0.10$ in
$S$ and $\pm 0.12$ in $T$ \cite{pdg}.  For plausible values of $\Lambda_3$, the
shifts in eqs. (\ref{scor}) and (\ref{tcor}) are quite safely smaller than
these uncertainties.  The corrections to $U$ are also negligibly small.

Thus, we find that this current precision electroweak data is consistent with
the power-law decays of running SM fermion masses, eq. (\ref{mflargep}), that
occur in theories with dynamical fermion mass generation.  Indeed, we find that
these effects are safely small compared with other contributions that can be
expected in such theories, such as technifermion loop contributions to $S$ in
TC and TC2 theories \cite{scalc,tc2}.  An interesting aspect of the corrections
(\ref{tcor}) and (\ref{scor}) is that even in what are ostensibly
standard-model inputs to these precision electroweak quantities there is
already a ``hidden'' effect of this new physics - specifically, the softness of
the SM fermion masses.

Soft standard-model fermion masses affect other one-loop processes where top
quarks make important SM contributions, such as $B_d -\bar B_d$ and $B_s - \bar
B_s$ mixing, and $Z \to b \bar b$ and $K^+ \to \pi^+ \nu\bar\nu$ decays.  Of
these, $BR(Z \to b \bar b)$ and the $\Delta m_{B_d}$ from $B_d -\bar B_d$
mixing are the most precisely measured, with fractional accuracies of $0.003$
and $0.01$, respectively \cite{pdg}. Soft top quark mass effects could be $
\sim {\rm few} \times 10^{-3}$ and are consistent with this data. Again, these
processes also receive (model-dependent) contributions from the new physics
\cite{vsmcsm}.  In ETC models, certain ETC gauge boson exchanges contribute to
the above processes, but do not couple to $W$ or $Z$ and hence do not directly
affect $S$ or $T$ at one-loop level.

Corrections due to soft SM fermion masses are also present in precisely
measured quantities involving first- and second-generation fermions.  However,
for the anomalous magnetic moments of the electron and muon, $a_e$ and $a_\mu$,
these corrections are expected to go like $m_e^2/\Lambda_1^2$ and
$m_\mu^2/\Lambda_2^2$, respectively. These are much smaller than the respective
fractional measurement accuracies of about $10^{-9}$ and $10^{-6}$ for $a_e$
and $a_\mu$ \cite{pdg,hk}. Specific dynamical models yield other corrections,
e.g., \cite{ckm}.

Finally, we note an important theoretical implication. Consider the one-loop
(primitively divergent) corrections to a SM fermion propagator.  These yield an
inverse fermion propagator $S_{f_j}(p)^{-1} = A_{f_j}(p)\pslash -
B_{f_j}(p)$.  In the SM the divergences in $ A_{f_j}$ and $B_{f_j}$ are
cancelled by wavefunction and mass renormalization so that, onshell,
$S_{f_j,ren.}(p)^{-1}=\pslash - m_{f_j}$. However, the power-law decay
(\ref{mflargep}) renders $B_{f_j}$ finite.  Thus, there is a change in what is
divergent versus what is finite, which means a change in the renormalization
procedure for the standard model viewed as the low-energy limit of a theory
with dynamical fermion mass generation. This is in accord with the greater
predictiveness of such theories.  

In summary, in asymptotically free theories that dynamically generate
SM fermion masses from underlying $\langle \bar\psi\psi\rangle$
condensate(s), we have shown that the running masses $m_{f_j}(p)$ have the
power-law decay (\ref{mflargep}) for $p \gg \Lambda_j$.  We have explored
effects of this and have shown that it is consistent with current precision
electroweak data.

We thank T. Appelquist, K. Lane, and M. Piai for helpful discussions and
comments. This research was partially supported by the grant NSF-PHY-00-98527.


\begin{thebibliography}{99}
\bibitem{eta} 
Here $\eta = \exp[\int_{\Lambda_{TC}}^{\Lambda_w}
\gamma(\alpha_{TC}(\mu)) d\mu/\mu]$, where $\gamma$ denotes the anomalous
dimension of $\bar F F$.  For walking TC with 
$\gamma \simeq 1$, this factor is $\eta \simeq \Lambda_w/\Lambda_{TC}$.
\bibitem{except} 
Note that eq. (\ref{mflargep}) would not apply to a hypothetical model
in which the SM fermion mass hierarchy arose from a hierarchy in the
condensate $({\rm mass})^3$ values themselves, with a (larger) fixed mass $M$
describing the vector bosons that connect SM fermions with the set of
$\{\psi\}$; such a model could plausibly yield $m_{f_j}(p) \sim m_{f_j}M^2/p^2$
for $p \gg M$.  The necessity of our asymptotic freedom assumption is 
illustrated by the fractional power law obtained in the non-asymptotically free
QED theory of Ref. \cite{jbw}.
\end{thebibliography}
\end{document}